\documentclass[aps,superscriptaddress,showpacs,pre]{revtex4}

\usepackage{amsmath,amssymb}
\usepackage{graphicx}

\def\t{\tilde}
\newcommand{\E}[1]{Eq.~(\ref{#1})}
\newcommand{\F}[1]{Fig.~\ref{fig:#1}}
\newcommand{\Sec}[1]{Sec.~\ref{sec:#1}}

\begin{document}

\title{slow switching in a population of delayed pulse-coupled oscillators}

\author{Hiroshi Kori}
\email{kori@ton.scphys.kyoto-u.ac.jp}
\affiliation{Department of Physics, Graduate School of Sciences, 
Kyoto University, Kyoto 606-8502, Japan}
\pacs{87.10.+e,05.45.Xt,05.90.+m}

\begin{abstract}   
 We show that peculiar collective dynamics called slow switching arises
 in a population of leaky integrate-and-fire oscillators with delayed,
 all-to-all pulse-couplings.  By considering the stability of cluster
 states and symmetry possessed by our model, we argue that saddle
 connections between a pair of the two-cluster states are formed under
 general conditions. Slow switching appears as a result of the system's
 approach to the saddle connections.  It is also argued that such saddle
 connections easy to arise near the bifurcation point where the state of
 perfect synchrony loses stability. We develop an asymptotic theory to
 reduce the model into a simpler form, with which an analytical study of
 cluster states becomes possible.
\end{abstract}

\maketitle

\section{introduction} \label{sec:introduction}
Studies on collective motion of coupled oscillators have attracted
considerable attention over the last few
decades\cite{winfree67,kuramoto84,synchronization}. It is commonly seen
that a population of autonomous elements performs certain biological
functions by behaving collectively\cite{fromneuron}. It has in fact
been pointed out that collective motion is crucial to information
processing and transmission in living organisms\cite{binding}.

In the brain, the neurons are exclusively coupled through chemical
synapses, i.e., the neurons communicate by pulses of
transmitter\cite{methods}.  Chemical synapses commonly form dense and
complex networks.  For mathematical modeling of neuronal networks,
homogeneous all-to-all (or, global) coupling is often adopted. Although
the global coupling may be a little too idealistic, the corresponding
networks share a lot of properties in common with systems with complex and
dense networks.

In the present paper, we consider a population of neural oscillators
with delayed, all-to-all pulse-coupling. The oscillator we use is called
the leaky integrate-and-fire (LIF) model. There are a large amount of
papers concerning LIF in physics and neuroscience, e.g., see
\cite{mirollo90,kuramoto91,sakai99}. This is because LIF is a quite
simple model still capturing some essential characteristics of neuronal
dynamics, i.e., it represents an integrator with relaxation, and resets
after it fires. Though our population model is commonly used, (e.g., see
\cite{abbott93}), its collective dynamics does not seem to have been
studied so carefully.  We are particularly concerned with peculiar
collective dynamics called slow switching\cite{hansel93,kori01}. The
study of collective dynamics in the original form of the model is not
easy to handle because the coupling involves a long term memory.  We
thus develop an asymptotic theory and reduce our model into a form
without memory, by which an analytical study of collective dynamics
becomes possible.

\section{model} \label{sec:model}
The population model we consider consists of $N$ identical elements with
delayed, all-to-all pulse-coupling. The dynamics of each elements is
described by a single variable $v_i$ ($i=1,2,\ldots,N$) which
corresponds to the membrane potential of a neuron. The equation for
$v_i$ is given by
\begin{equation} 
 \label{model}
 \frac{d}{dt}v_i(t) = a - v_i +  \frac{K}{N} (b-v_i) E(t) .
\end{equation}
The parameter $a$ is the so-called resting potential to which $v_i$
relaxes when the coupling is absent.  It is assumed that when $v_i$
reaches a threshold value which is set to $1$, it is instantaneously
reset to zero. This event is interpreted as a spiking event. The
dynamics is thus called LIF.  When a neuron spikes, it emits a pulse
toward each neuron coupled to it, and the latter receives the
pulse with some delay called a synaptic delay. The coupling is assumed
to be homogeneous and all-to-all, so that its effect can be represented
by one global variable $E$, given by
\begin{equation}
 \label{E}
 E(t)= \sum_{j=1}^N \sum_{\rm spikes} e(t-t_j^{\rm spikes}-\tau).
\end{equation}
Here, ${t_j^{\rm spikes}}$ represents a series of times at which the
$j$-th neuron spikes and $\sum_{\rm spikes}$ denotes a summation over
the series of such spikes; $\tau$ is the synaptic delay, and $e(t)$ is a
{\em pulse function}, given by
\begin{equation} 
\label{e}
 e(t)= \frac{\alpha \beta}{\beta-\alpha}
  (e^{-\alpha t}- e^{-\beta t}) \Theta(t),
\end{equation}
where $\Theta(t)$ is the Heaviside function; $\alpha$ and $\beta$ are
constants satisfying $\beta>\alpha$.  In the limit $\beta\to\alpha$,
$e(t)$ becomes $\alpha^2 t e^{-\alpha t}$, which is called the alpha
function\cite{methods}.  $b$ is called the reversal potential to which
$v_i$ relaxes when $E(t)$ is positive, i.e., while the neuron receives
the pulses. $K$ is a positive constant characterizing the strength of
the coupling.  The coupling assumed above is characteristic to the
synaptic coupling.  The coupling become excitatory (EPSP) if $b>1$, and
inhibitory (IPSP) if $b \le 0$.

If $a\le1$, LIF becomes an excitable neuron, while if $a>1$, it repeats
periodic spikes, namely, it represents a neural oscillator. We assume
$a>1$ throughout the present paper, and call each element an {\em
oscillator}. Then, we can define a variable $\psi_i$ varying smoothly
with time, which turns out useful in the following discussion. We call
$\psi_i$ the {\em phase} of the $i$-th oscillator, and define it by
\begin{equation} 
 \label{phase}
  \psi_i = \int_0^{v_i} \frac{dv}{a-v} =\ln\left(\frac{a}{a-v_i}\right),
\end{equation}
which varies between $0$ and the intrinsic period of oscillation $T$
given by
\begin{equation} 
 \label{period}
 T = \ln \left( \frac{a}{a-1} \right).
\end{equation}
Note that $\psi_i$ satisfies $d\psi_i/dt=1$ in the absence of coupling.

\section{numerical results} \label{sec:results}
By numerically integrating our model under random initial distributions
of $v_i$, we find various types of collective behavior. Among them, we
are particularly interested in the slow switching phenomenon, which can
arise when $b>a$ and $N\ge4$. As displayed in \F{slowswitching}, the
whole population, which was initially distributed almost uniformly,
splits into two subpopulations, each of which converges almost to a
point cluster. However, after some time the phase-advanced cluster
starts to scatter. Then, this scattered group starts to converge again
as it comes behind the preexisting cluster. In this way, the preexisting
cluster becomes a phase-advanced cluster.  After some time, again, this
phase-advanced cluster begins to scatter, and a similar process repeats
again and again. In other words, the system switches back and forth
between a pair of two-cluster states. For larger times, the system comes
closer to each of well-defined two-cluster states and stays near the
state longer. Theoretically, these switchings repeats indefinitely,
although in numerical integrations the system converges at one of the
two-cluster states in a finite time and stops switching due to numerical
round-off errors\cite{kori01}.

The slow switching phenomenon occurs within a broad range of parameter
values provided that $K$ is small, and the time constants
$\alpha^{-1},\beta^{-1}$ and $\tau$ are small compared with $T$. For
larger $\alpha^{-1},\beta^{-1}$ and $\tau$, the slow switching
phenomenon becomes less probable, and the appearance of steady
multi-cluster states becomes more probable instead. For $b<a$, we find
no two-cluster states involving slow switching, while steady
multi-cluster states are observed in most cases. The corresponding phase
diagram will be presented in \Sec{hetero} (see \F{stability}).

\section{weak coupling limit} \label{sec:asymptotic}
Our model given by \E{model} is relatively simple, still it would be
difficult to get some insight into its collective dynamics
analytically. Fortunately, however, our main results given in
\Sec{results} do not change qualitatively in the weak coupling limit,
i.e., $K\to0$.  In this limit, our model is reduced to a much simpler
form with which we can study the existence and stability of various
cluster states analytically. Derivation of the reduced model is given as
follows.

\begin{figure}
 \includegraphics{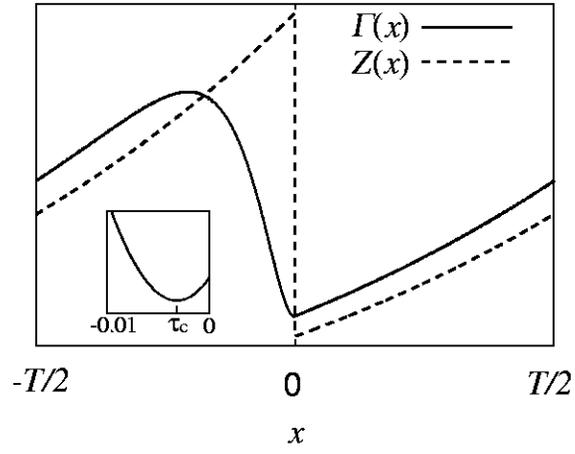} \caption{The solid line shows the coupling
 function $\Gamma(x)$ for $a=1.05$ ($T\simeq 3.0$), $\alpha^{-1}=0.2$
 and $\beta\to\alpha$. The minimum appears at $x=\tau_c$ which is a
 small negative. For comparison, the shape of $Z(x)$ is also displayed
 with the broken line.}  \label{fig:gamma}
\end{figure}

Substituting $v_i = a(1-e^{-\psi_i})$ into Eq. (\ref{model}), we obtain
\begin{equation}
 \label{gomi2}
 \frac{d}{dt}\psi_i(t) = 1 + \frac{K}{N} \sum_{j=1}^N \sum_{\rm spikes}
  Z(\psi_i)e(t-t_j^{\rm spikes}-\tau),
\end{equation}
where
\begin{equation}
 \label{Z}
 Z(\psi_i)   = \frac{b-a}{a} e^{\psi_i}+ 1.
\end{equation}
It is convenient in the following calculation to redefine $Z$ as a
$T$-periodic function, or, $Z(\psi_i+nT)=Z(\psi_i)$ ($n=\pm 1,\pm
2,\cdots$). Note that sudden drop of $Z(x)$ at $x=0$ is due to our rule
employed, i.e., the membrane potential is instantaneously reset at
$v_i=1$.  We also define a residual phase $\Psi_i$ by
\begin{equation}
 \label{Psi}
 \Psi_i = \psi_i -t.
\end{equation}
Substituting \E{Psi} into \E{gomi2}, we obtain
\begin{equation} 
 \label{model2}
 \frac{d}{dt}\Psi_i(t) =  \frac{K}{N} \sum_{j=1}^N \sum_{\rm spikes}
  Z ( \Psi_i+t) e(t-t_j^{\rm spikes}-\tau).
\end{equation}
We now assume that $K$ is sufficiently small so that the r.h.s of
Eq.~(\ref{model2}) is sufficiently smaller than the intrinsic frequency
$T^{-1}$. This allow us to make averaging of the r.h.s of
Eq.~(\ref{model2}) over the period $T$. The zeroth order approximation
with respect to the smallness of $K$, which corresponds to the free
oscillations, is given by
\begin{equation} 
 \label{kinji1}
 \Psi_i(t) = \rm const.
\end{equation}
and
\begin{equation} 
 \label{kinji2}
 t_j^{\rm spikes} = t_j - nT, \quad (n=0,1,2,\ldots),
\end{equation}
where $t_j$ is the latest time at which the $j$-th neuron spikes.  In the
first order approximation, we may time-average Eq.~(\ref{model2}) over
the range between $t-T$ and $t$ using Eqs.~(\ref{kinji1}) and
(\ref{kinji2}):
\begin{eqnarray}
 \label{averaging}
 \frac{d}{dt}\Psi_i(t) &=& \frac{K}{N} \sum_{j=1}^N  \frac{1}{T}
  \int_{t-T}^{t} \sum_{n=0}^{\infty}
  Z ( \Psi_i(t)+\lambda) e(\lambda-t_j+nT-\tau) d\lambda \\
 \label{tatami}
  &=& \frac{K}{N} \sum_{j=1}^N  \frac{1}{T}
  \int_{0}^{\infty} Z ( \Psi_i(t)+t_j+\tau+\lambda') e(\lambda') d\lambda' \\
 \label{pm2} 
  &=& \frac{K}{T} + \frac{K(b-a)}{Na} 
  \sum_{j=1}^N \Gamma \bigr( \Psi_i(t)+t_j+\tau \bigl),
\end{eqnarray}
where
\begin{equation}
 \label{gamma}
  \Gamma(x) = \frac{\alpha\beta}  {\beta-\alpha}
  \bigl\{ H_{\alpha,T}(x) - H_{\beta,T}(x) \bigr\},
\end{equation}
\begin{eqnarray} 
 \label{H}
 H_{\alpha,T}(x) &=& \frac{1}{T} 
  \int_0^{\infty}  \exp[{ (x+\lambda)\ {\rm mod} \ T}]
  \exp[{-\alpha \lambda}] d\lambda \nonumber \\
  &=& \frac{(e^T-1)\exp[{\alpha (x\  {\rm mod} \ T)}]-(e^{\alpha T}-1)\exp[{x\ {\rm mod} \ T}] } 
  {T(1-\alpha)(e^{\alpha T}-1)}.
\end{eqnarray}
Note that $\Gamma(x)$ and $H_{\alpha,T}(x)$ are $T$-periodic functions.
Figure \ref{fig:gamma} illustrates a typical shape of the coupling function
given by \E{gamma}.  Furthermore, using the identity
\begin{equation}
 \Psi_j(t_j) = \psi_j(t_j)-t_j = -t_j,
\end{equation}
and the zeroth order approximation $\Psi_j(t_j)=\Psi_j(t)$, we may
replace $t_j$ by $-\Psi_j(t)$ in \E{pm2} in the first order
approximation. Thus, we finally obtain
\begin{equation}
 \label{pm} 
 \frac{d}{dt}\psi_i(t) = \omega + \frac{K'}{N} 
  \sum_{j=1}^N \Gamma \bigr( \psi_i(t)-\psi_j(t)+\tau \bigl),
\end{equation}
where $\omega=1+K/T$ and $K'=K(b-a)/a$. Equation (\ref{pm}) is the
standard form of the phase model. Note that the error involved in
\E{kinji2} may look to diverge as $n\to\infty$, still the final error
vanishes in the first order approximation due to the decay of $e(t)$. It
should be noted that the reduced model is free from memory effects, but
the effect of delay has been converted to a phase shift in the coupling
function. Similar form of the phase model is generally obtained in
delayed coupled oscillators when the coupling is sufficiently
weak\cite{kori01}. Hereafter, we ignore the degree of freedom associated
with the dynamics of the center of mass (or, mean phase) which can be
decoupled in the phase model.

Important parameters of our phase model given by \E{pm} with \E{gamma}
are $T,\alpha,\beta,\tau$ and the sign of $K'$ (i.e., the sign of
$b-a$). The reason is the following. We may take $|K'|=1$ by properly
rescaling of $t$ and $\omega$, while its sign is crucial because the
local stability of any solution depends on it. $\omega$ gives the
frequency of steady rotation of the whole system, which is irrelevant to
collective dynamics.  We choose $T$ as an independent parameter by which
$a$ becomes dependent through \E{period}.  It is remarkable that our
coupling function is independent of $b$.  In fact, change in $b$ causes
no qualitative change in our result as far as the sign of $b-a$ remains
the same.  Interestingly, even if we replace the term $b-v_i$ by a
constant $c$ in \E{model}, i.e.,
\begin{equation} 
 \label{model3}
 \frac{d}{dt}v_i(t) = a - v_i +  \frac{Kc}{N} E(t),
\end{equation}
then we can reduce this model similarly and obtain the same coupling
function as in Eq.(\ref{gamma}). We have checked that \E{model3}
actually reproduces qualitatively the same results as those given in
\Sec{results}. In that case, negative $c$ corresponds to the case $b<a$
in \E{model}.

In the following section, we assume $b>a$ and $\beta\to\alpha$ unless
stated otherwise.

\section{two-oscillator system} \label{sec:two-oscillator}
\begin{figure}
 \includegraphics{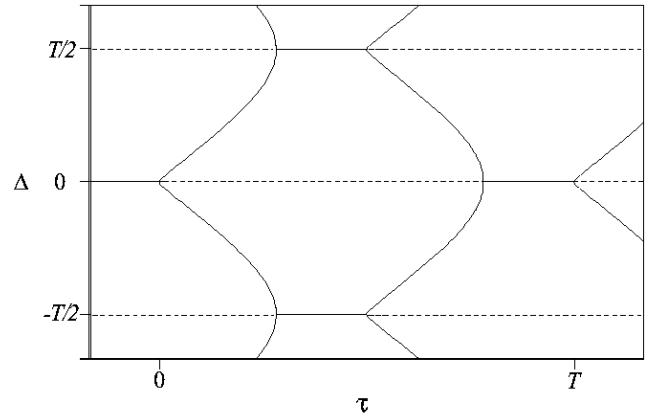} \caption{Bifurcation diagram of a two-oscillator
 system. Solid and dotted lines respectively represent stable and
 unstable branches, where $b>a$ is assumed. The stability property is
 reversed if $b<a$.}  \label{fig:diagram}
\end{figure}

In this section, we study a two-oscillator system, or, $N=2$. Although
the two-oscillator system is not directly related to the main subject of
the present paper, one may learn some basic properties of our phase
model from this simple case. Defining $\Delta=\psi_1-\psi_2$, we obtain
\begin{equation}
 \label{n2-x}
  \frac{d\Delta}{dt} = \frac{K'}{2}\bigl( \Gamma(\Delta+\tau)
  -\Gamma(-\Delta+\tau) \bigr)
  \equiv G_{\tau}(\Delta).
\end{equation}
Phase locking solutions are obtained by putting $d\Delta/dt = 0$, and
the associated eigenvalues are given by $dG_{\tau}/d\Delta$. Figure
\ref{fig:diagram} shows a bifurcation diagram of the phase locking
solutions, in which we take $\tau$ as a control parameter. We find that
for small $\tau$ the trivial solutions $\Delta=0$ (in-phase locking) and
$T/2$ (anti-phase locking) are unstable, while there are a pair of
stable branches of non-trivial solutions. The point $\tau=0$ is close to
the bifurcation point where the in-phase state loses stability. The
bifurcation occurs at $\tau=\tau_{\rm c}$, where $\tau_{\rm c}$
corresponds to the minimum of $\Gamma(x)$ (see \F{gamma}).  Because
$\tau_{\rm c}$ is negative, the in-phase state cannot be stable for
small or vanishing delays (while it can be stable for delays comparable
to $T$ due to the $T$-periodic nature of our phase model). $\tau_{\rm
c}$ is extremely small, which is due to the sudden drop of $Z(x)$ at
$x=0$ and the particular rule employed in our model, i.e., a neuron is
assumed to spike and reset simultaneously. The width of the stable
branches of the trivial solutions is the same as that of the decreasing
part of $\Gamma(x)$.  Owing to the peculiar shape of $Z(x)$, the width
is of the same order as the width of $e(t)$, which is $O(\alpha^{-1})$.
The stability of the in-phase state is identical with that of the state
of perfect synchrony.

In terms of the original model, we now present a qualitative
interpretation of why the in-phase locking state is unstable for small
or vanishing delays.  We consider the dynamics of two neurons which are
initially very close in phase. The effect of a pulse on the phase
$\psi_i$ is larger for smaller $dv_i/dt$.  $dv_i/dt$ is monotonously
decreasing except when it is reset (which reflects on the property of
$Z(x)$ that it is increasing except $x=0$).  Thus, the neuron with
larger $v_i$ makes a larger jump in phase when it receives a pulse, by
which the phase difference between the two neurons becomes larger when
they receive a pulse.  On the other hand, the situation becomes
different if two neurons lie before and after the resetting point, i.e.,
if the phase-advanced neuron has smaller $v_i$. In that case, the phase
difference becomes smaller when they receive a pulse. According to our
dynamical rule, however, resetting and spiking occur simultaneously, so
that they receive pulses when the phase-advanced neuron has larger
$v_i$. Therefore, the in-phase state becomes inevitably unstable even
without delay. If we want to obtain a stable in-phase state for small
delays, we should employ a rule such that a neuron spikes before it is
reset, which would be more physiologically plausible than the rule
employed here.

\section{local stability analysis for a large population}\label{sec:stability}
The trivial in-phase solution and the non-trivial solutions of the
two-oscillator system correspond respectively to the state of perfect
synchrony and two-cluster states when we go over to a large population.  In
this section, we study local stability of the two-cluster
states. Although non-trivial solutions are stable for small or vanishing
delays in the two-oscillator system, the corresponding two-cluster
states turn out unstable.

We consider a steadily oscillating two-cluster state in which the two
clusters consist of $Np$ and $N(1-p)$ oscillators, respectively.  The
oscillators inside each cluster are completely phase-synchronized, and
the phase-difference between the clusters is constant in time, which is
denoted by $\Delta$.  From \E{pm}, the phase difference obeys the equation
\begin{equation}
 \label{dtdelta}
 \frac{d\Delta}{dt}=K' \left\{(2p-1)\Gamma(\tau)+
 (1-p)\Gamma(\Delta+\tau)+p\Gamma(-\Delta+\tau) \right\}.
\end{equation}
When $\Delta$ is constant, we obtain a relation between $p$ and
$\Delta$ as
\begin{equation}
 \label{p-delta}
 (2p-1)\Gamma(\tau)+ (1-p)\Gamma(\Delta+\tau)+p\Gamma(-\Delta+\tau) =0.
\end{equation}
We designate a two-cluster state satisfying \E{p-delta} as $(p,\Delta)$.
The eigenvalues of the stability matrix are calculated as
\begin{equation}
 \label{l1}
  \lambda_{1} = K'\{p\Gamma'(\tau)+(1-p)\Gamma'(\Delta+\tau)\}, 
\end{equation}
\begin{equation}
 \label{l2}
  \lambda_{2} = K'\{(1-p)\Gamma'(\tau)+p\Gamma'(-\Delta+\tau)\},  
\end{equation}
\begin{equation} 
 \label{l3}
  \lambda_{3} = K'\{(1-p)\Gamma'(\Delta+\tau)+p\Gamma'(-\Delta+\tau)\},
\end{equation}
where $\Gamma'(x)$ means $(d/dx)\Gamma(x)$.  The multiplicities of
$\lambda_1,\lambda_2$ and $\lambda_3$ are $Np-1$,$N(1-p)-1$ and $1$,
respectively. $\lambda_1$ and $\lambda_2$ correspond to fluctuations in
phase of the two oscillators inside the phase-advanced and phase-retarded
cluster, respectively. $\lambda_{3}$ corresponds to fluctuations in the
phase difference $\Delta$ between the clusters.

\begin{figure*}
 \includegraphics{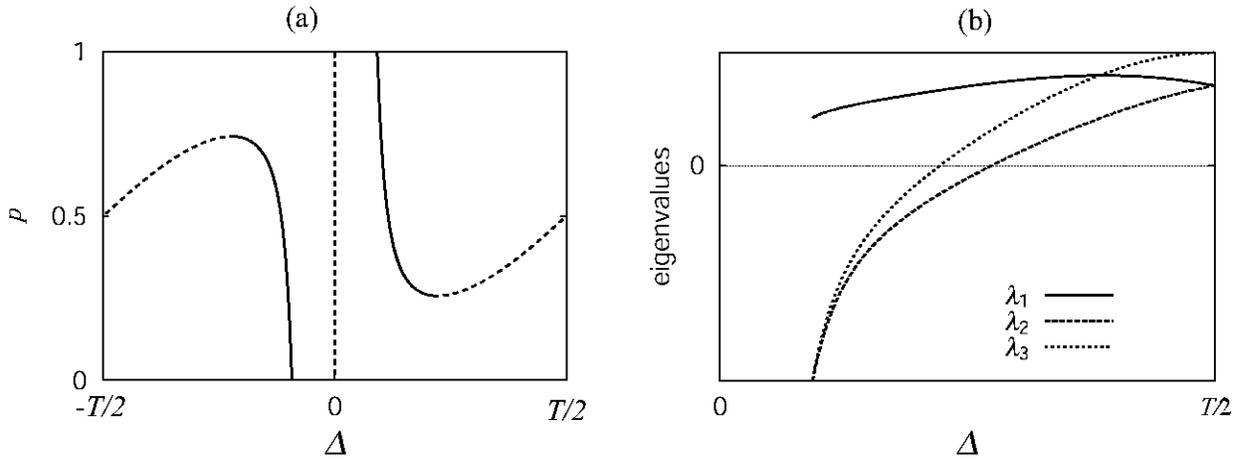} \caption{(a)Relation between $p$ and $\Delta$
 associated with two-cluster states. (b)Eigenvalues of two-cluster
 states as a function of $\Delta$. In (a), the solid and dotted lines
 correspond to the two-cluster state of negative and positive
 $\lambda_3$, respectively.}  \label{fig:p-delta}
\end{figure*}

Substituting \E{gamma} into \E{p-delta}, we obtain a relation between
$p$ and $\Delta$, the corresponding curve being shown in
\F{p-delta}(a). By using this relation, the eigenvalues of $(p,\Delta)$
can be obtained, which are displayed in \F{p-delta}(b) as a function of
$\Delta$. It is found that no two-cluster states are stable, and there
is a set of $(p,\Delta)$ for which $\lambda_1>0$ and
$\lambda_2,\lambda_3<0$.  Positive $\lambda_1$ means that the
two-cluster state is unstable with respect to perturbations inside a
phase-advanced cluster. On the other hand, $\lambda_2,\lambda_3<0$ means
that it is {\em stable} against perturbations inside a phase-retarded
cluster as far as the perfect phase-synchrony of the phase-advanced
cluster is maintained. Within a certain range of $p$, there are pairs of
two-cluster states represented by $(p,\Delta)$ and $(p,\Delta')$ with
the same stability property, and they appear as the solid lines in
Fig~\ref{fig:p-delta}(a). In the next section, we explain how a {\em
heteroclinic loop} between the $(p,\Delta)$ and $(p,\Delta')$ is
stably formed in our model.

Similarly to the discussion in \Sec{two-oscillator}, the stability
property mentioned above can also be understood in terms of the original
model.  Every neuron inside the phase-advanced cluster always receives
pulses when its membrane potential is increasing. Then, the
phase-difference between two neurons inside the cluster, if any, always
increases, so that the phase-advanced cluster
is inevitably unstable. On the other hand, the neurons inside the
phase-retarded cluster can receive pulses (emitted by the
phase-advanced cluster) during their resetting. Then, the
phase-differences between neurons inside the phase-retarded cluster, if
any, become smaller, so that the phase-retarded cluster can be stable.

\section{heteroclinic loop} \label{sec:hetero}
We first note that there is a particular symmetry of our
model which turns out crucial to the persistent formation of the
heteroclinic loop. The symmetry is given by
\begin{equation} 
 \label{identity_v}
 \left. \frac{d}{dt} \{v_i(t)-v_j(t) \} 
 \right|_{v_i(t)=v_j(t)}=0 \quad \mbox{for any $i$ and $j$}.
\end{equation}
Due to this symmetry, the units which have the same membrane potential at
a given time behave identically thereafter. In other words, once a point
cluster is formed, it remain a point cluster forever.

We assume that a pair of two-cluster states (called A and B) exists and
has the same stability property as that discussed in \Sec{stability},
i.e., the phase-advanced cluster is unstable, and the phase-retarded
clusters is stable. Suppose that our system is in a two-cluster state A
initially. When the oscillators inside the phase-advanced cluster are
perturbed while the phase-retarded cluster is kept unperturbed [see
\F{zukai}(a)], the former begins to disintegrate while the latter
remains a point cluster. Then, the group of dispersed oscillators and the
point cluster coexist in the system [see \F{zukai}(b)].  We know,
however, that in the presence of a point cluster, there exists a stable
two-cluster state in which the existing point cluster is
phase-advanced. From this fact, the dispersed oscillators are expected
to converge to form a point cluster coming behind the preexisting point
cluster.  In this way, the system relaxes to another two-cluster state B
[see \F{zukai}(c)]. From the above statement, it is implied that in our
high-dimensional phase space there exists a saddle connection from the
state A to the state B.  The existence of a saddle connection from the
state B to the state A can be argued similarly.  A heteroclinic loop is
thus formed between the pair of the two cluster states A and B.

\begin{figure}
 \includegraphics{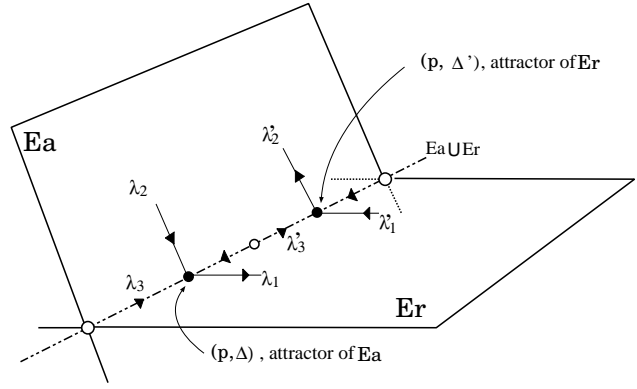} \caption{Schematic representation of the
 structure of a heteroclinic loop. $(p,\Delta)$ and $(p,\Delta')$ become
 attractors in the invariant subspaces $E_a$ and $E_r$, respectively.}
 \label{fig:hetero}
\end{figure}

In terms of the phase model, the above argument can be reinterpreted in
a little more precise language. In the phase model given by
Eq.~(\ref{pm}), a symmetry property similar to \E{identity_v} also
holds:
\begin{equation} 
 \label{identity}
 \left. \frac{d}{dt} \{\psi_i(t)-\psi_j(t) \} 
 \right|_{\psi_i(t)=\psi_j(t)}=0 \quad \mbox{for any $i$ and $j$}.
\end{equation}
Our argument will be based on the following assumptions:
\begin{itemize}
 \item[(a)] $(p,\Delta)$ with $\lambda_1>0$ and $\lambda_2,\lambda_3<0$ exits,
 \item[(b)] $(p,\Delta')$ with $\lambda'_2>0$ and $\lambda'_1,\lambda'_3<0$ exits,
\end{itemize}
where we define $\Delta>0$ and $\Delta'<0$, and $\lambda_i$ and
$\lambda'_i$ ($i=1,2,3$) are the eigenvalues of $(p,\Delta)$ and
$(p,\Delta')$, respectively. Note that if $p=0.5$, the two clusters in
question are identical, or $\Delta'=\Delta$, so that (a) and (b) are
identical. Figure \ref{fig:hetero} illustrates a schematic presentation
of the $N-1$ dimensional phase space structure, where we ignore the
degree of freedom associated with the dynamics of the center of
mass. $E_a$ and $E_r$ are identical with the subspaces where the
phase-advance and phase-retarded clusters of $(p,\Delta)$ remain point
clusters, respectively. By considering the direction of eigenvectors,
one can easily confirm that $E_a$ and $E_r$ are identical with the
stable subspaces of $(p,\Delta)$ and $(p,\Delta')$,
respectively. Furthermore, because $E_a$ and $E_r$ are {\em invariant}
subspaces due to the symmetry given by Eq.~(\ref{identity}),
$(p,\Delta)$ and $(p,\Delta')$ are attractors within $E_a$ and $E_r$,
respectively. Thus, a heteroclinic loop between $(p,\Delta)$ and
$(p,\Delta')$ should necessarily exits. The saddle connections in
question are stably formed through the invariant subspaces which exist
for the symmetry of equations of motion given by Eq.~(\ref{identity}). The
heteroclinic loop is thus robust against small perturbations to the system
unless the symmetry is broken.

Whether the resulting heteroclinic loop is attracting or not depends on
the following quantity:
\begin{equation}
 \label{exponent}
 \gamma \equiv \frac{\lambda_2\lambda'_1}{\lambda_1\lambda'_2}.
\end{equation}
It was argued in Ref.~\cite{hansel93} that if $\gamma>1$, the system can
approach the heteroclinic loop and come to move along it. In that case,
the time interval during which the system is trapped to in the vicinity
of one of the two-cluster states increases exponentially with time.
Substituting the eigenvalues obtained from Eqs.~(\ref{l1}) and
(\ref{l2}) using Eq.~(\ref{gamma}) into Eq.~(\ref{exponent}), we find
that the heteroclinic loops within a certain range of $p$ are in fact
attracting for small $\alpha^{-1},\beta^{-1}$ and $\tau$. Phase diagrams
of the heteroclinic loops and symmetric multi-cluster states is shown in
\F{stability}, where we choose $\tau$ as a control parameter (see
appendix for the stability of the symmetric multi-cluster states).

\section{near the bifurcation point}\label{sec:near}
In this section, we concentrate on the vicinity of the bifurcation point
where the state of perfect synchrony loses stability. As noted in
\Sec{two-oscillator}, the bifurcation occurs at $\tau=\tau_{\rm c}$.
Then, for small $x-\tau_{\rm c}$, the coupling function can be expanded as
\begin{equation}
 \label{gamma-appro}
 \Gamma(x)=c_0+c_2(x-\tau_{\rm c})^2-c_3(x-\tau_{\rm c})^3+O((x-\tau_{\rm c})^4).
\end{equation}
Suppose that $c_2$ and $c_3$ are positive. We further put $c_3=1$ by
properly rescaling $K'$ in \E{pm}. In order to find possible
two-cluster states, we solve \E{p-delta} using \E{gamma-appro}. We then
obtain three solutions for $\Delta$ as a function of $p$ and $\tau$.
One is the trivial solution $\Delta=0$ (the perfect synchrony), and the
others are given by
\begin{equation}
 \label{delta}
 \Delta = 
  \frac{(1-2p)(c_2-3\t\tau)}{2} \pm \sqrt{\frac{(1-2p)^2(c_2-3\t\tau)^2}{4}+2c_2\t\tau},
\end{equation}
where $\t\tau\equiv\tau-\tau_{\rm c}$.  Note that the expression above
using the approximate $\Gamma$ given by \E{gamma-appro} is valid only
for small $\Delta$, which is actually the case if $p$ is close to $1/2$
and $\t\tau$ is small.  Substituting the expressions in \E{delta} into
Eqs.~(\ref{l1})-(\ref{l3}), we obtain eigenvalues associated with the
two-cluster states.  The resulting bifurcation diagram for given $p$ is
shown in \F{bunki}. The solid and broken lines give the branches of
negative and positive $\lambda_3$, respectively.  Two solid branches
exist for $\tau>0$, which are represented by $(p,\Delta)$ and
$(p,\Delta')$ with $\Delta>0$ and $\Delta'<0$. One can easily confirm
that the eigenvalues of these states satisfy $\lambda_1,\lambda'_2>0$
and $\lambda_2,\lambda_3,\lambda'_1,\lambda'_3<0$ for arbitrary $p$ and
small $\t\tau$, which agree with the condition for the existence of a
heteroclinic loop. The quantity $\gamma$ defined by \E{exponent} can
also be calculated and turns out to be larger than $1$. Thus, all the
local stability conditions for the existence of an attracting
heteroclinic loop are generally satisfied just above the bifurcation
point provided $c_3>0$.

It is also possible that a heteroclinic loop is formed when $c_3<0$.  In
that case, it is expected to arise {\em subcritically}, so that both the
heteroclinic loop and the state of perfect synchrony may be stable over
some region of negative $\tilde\tau$.  In fact, we found that such
bistability arises in a population of the Morris-Leccar
oscillators\cite{morris81} with the same coupling form as in \E{model},
and an analysis by means of the phase dynamics actually shows that $c_3$
is negative.  To confirm the corresponding bifurcation structure, we
have to consider higher orders of $x-\tau_{\rm c}$ in the coupling
function. The details of this issue are omitted here.

\section{conclusions and discussion} 
We have discussed the slow switching phenomenon in a population of
delayed pulse-coupled oscillators. We found that the phenomenon is
caused by the formation of an attracting heteroclinic loop between a
pair of two-cluster states.  A particular stability property of the
two-cluster states and a certain symmetry of our model are responsible
for its formation.  Our original model given by \E{model} is reduced to
the standard phase model in the weak coupling limit, by which we
succeeded in studying the stability of the two-cluster states
analytically, and confirming the structure of the heteroclinic loop.  It
was also argued that under the mild condition of the coupling function
all the local stability conditions for the existence of an attracting
heteroclinic loop are generally satisfied just above the bifurcation
point.

The physical mechanism of the formation of a heteroclinic loop we
describe in \Sec{hetero} does not depend on the nature of elements
(e.g., phase oscillator, limit-cycle oscillator, excitable elements,
chaotic elements) and couplings (e.g., diffusive coupling, pulse
coupling). It is expected, therefore, that a heteroclinic loop arises in
a wide class of models of coupled elements.

\begin{acknowledgments}
 The author thanks Y.~Kuramoto for a critical reading of the manuscript.
 He also thanks T.~Aoyagi, T.~Chawanya, H.~Sakai and H.~Nakao for
 fruitful discussions.
\end{acknowledgments}

\appendix
\section{}
According to Ref.\cite{okuda93}, we summarize here the existence and the
stability analysis of {\em symmetric multi-cluster states} in the phase
model given by \E{pm}. In the symmetric $n$-cluster state, it is assumed
that each cluster consists of $N/n$ oscillators. We denote the phase of
cluster $k$ as $\phi_k$ ($k=0,1,\ldots,n-1$). There always exists the
following solution:
\begin{equation}
 \phi_k = \Omega t + \frac{Tk}{n},
\end{equation}
with
\begin{equation}
 \Omega = \frac{K'}{n}\sum_{k=0}^{n-1}\Gamma\left(\frac{Tk}{n}+\tau\right),
\end{equation}
which corresponds to the state in which the $n$ clusters are equally
separated in phase and rotate at a constant frequency $\Omega$.  The associated
eigenvalues are calculated as
\begin{equation}
 \lambda_{\rm intra}=\frac{K'}{n}\sum_{k=0}^{n-1}
  \Gamma'\left( \frac{Tk}{n} +\tau \right),
\end{equation}
\begin{equation}
 \lambda_{\rm inter}^p=\frac{K'}{n}\sum_{k=0}^{n-1}
  \Gamma'\left( \frac{Tk}{n} +\tau \right)
  (1-\exp[-i Tkp/n]).
\end{equation}
$\lambda_{\rm intra}$ is a intra-cluster eigenvalue with multiplicity of
$N-n$. $\lambda_{\rm inter}^p$ (p=1,\ldots,n-1) are associated with
inter-cluster fluctuations. If all of these eigenvalues have negative real
part, the symmetric $n$-cluster state is stable.

\begin{figure}
 \includegraphics{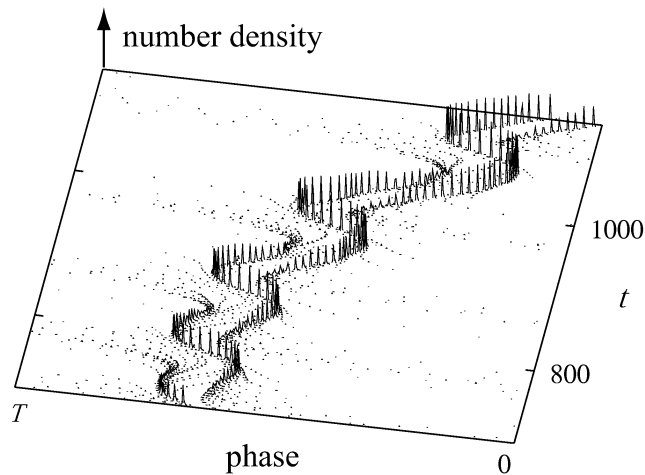} \caption{Slow switching phenomenon viewed
 through the number density of the oscillators as a function of
 phase. In order to get a better view, we work with a comoving frame of
 reference. The parameter values are $a=1.03$ ($T\simeq 3.5$), $b=2.0,
 \alpha^{-1}=0.3,\beta\to\alpha,\tau=0.2,K=0.1$ and $N=100$.}
 \label{fig:slowswitching}
\end{figure}
%
%
%
%
\begin{figure*}
 \includegraphics{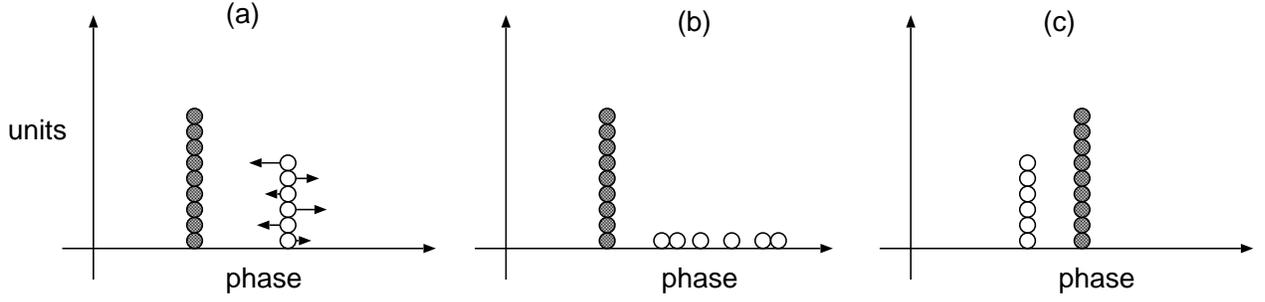} \caption{Schematic representation of a saddle
 connection between a pair of two cluster states, starting with the
 two-cluster state A (a) ending up with the other two-cluster state B
 (c).}  \label{fig:zukai}
\end{figure*}
%
%
\begin{figure*}
 \includegraphics{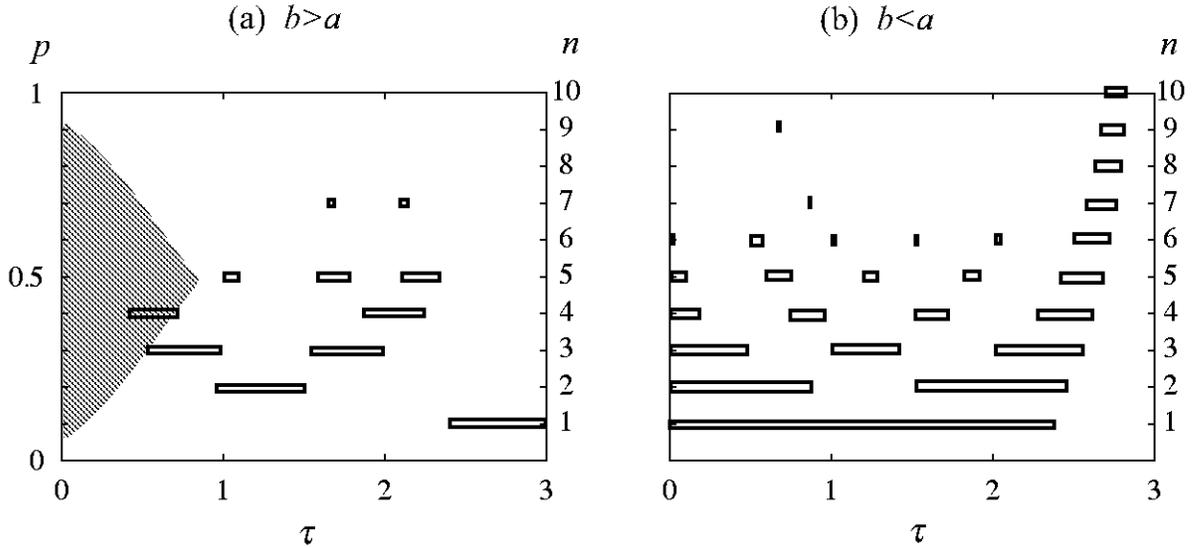} \caption{Phase diagrams of cluster states, where
 $\tau$ is chosen as a control parameter. The parameter values are the
 same as in \F{gamma} with (a)$b>a$ and (b)$b<a$, respectively. For
 given $p$ and $\tau$ inside the gray region, $\gamma$ is larger than
 one, i.e., the heteroclinic loop between $(p,\Delta)$ and
 $(p,-\Delta')$ is attracting. Each rectangle placed at $n$ indicates
 the region of $\tau$ within which the symmetric $n$-cluster state is
 stable. In (b), stable symmetric $n$-cluster states with $n>10$ also
 exist (not shown).}  \label{fig:stability}
\end{figure*}
%
\begin{figure}
 \includegraphics{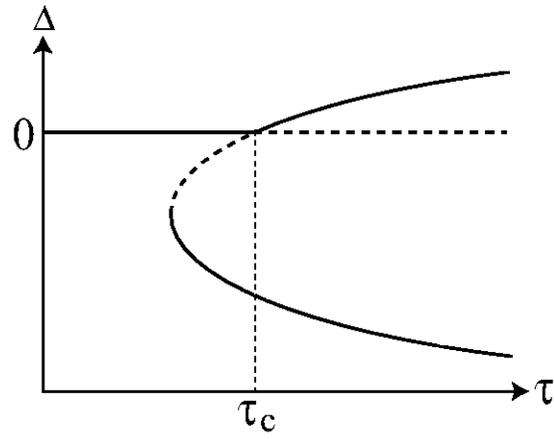} \caption{Bifurcation diagram around
 $\tau=\tau_c$. A heteroclinic loop is formed between a pair of the
 solid branches for $\tau>\tau_c$.}  \label{fig:bunki}
\end{figure}
%

\end{document}